\documentclass[11pt]{article} 
\usepackage{amsfonts,amsmath,amsxtra}
\usepackage{latexsym}
\usepackage{amssymb}

\def\hybrid{\topmargin 0pt      \oddsidemargin 0pt
        \headheight 0pt \headsep 0pt
        \textwidth 16.5cm
        \textheight 23cm
        \marginparwidth 0.0in
        \parskip 5pt plus 1pt   \jot = 1.5ex}
\catcode`\@=11
\def\marginnote#1{}
\newcount\hour
\newcount\minute
\newtoks\amorpm
\hour=\time\divide\hour by60 \minute=\time{\multiply\hour by60
\global\advance\minute by-\hour}
\edef\standardtime{{\ifnum\hour<12 \global\amorpm={am}%
        \else\global\amorpm={pm}\advance\hour by-12 \fi
        \ifnum\hour=0 \hour=12 \fi
      \number\hour:\ifnum\minute<10 0\fi\number\minute\the\amorpm}}
\edef\militarytime{\number\hour:\ifnum\minute<10 0\fi\number\minute}

\def\draftlabel#1{{\@bsphack\if@filesw {\let\thepage\relax
   \xdef\@gtempa{\write\@auxout{\string
      \newlabel{#1}{{\@currentlabel}{\thepage}}}}}\@gtempa
   \if@nobreak \ifvmode\nobreak\fi\fi\fi\@esphack}
        \gdef\@eqnlabel{#1}}
\def\@eqnlabel{}
\def\@vacuum{}
\def\draftmarginnote#1{\marginpar{\raggedright\scriptsize\tt#1}}

\def\draft{\oddsidemargin -0.1truein
        \def\@oddfoot{\sl preliminary draft \hfil
        \rm\thepage\hfil\sl\today\quad\militarytime}
        \let\@evenfoot\@oddfoot \overfullrule 3pt
        \let\label=\draftlabel
        \let\marginnote=\draftmarginnote
\def\@eqnnum{{\rm (\theequation)}
\rlap{\kern\marginparsep\tt\@eqnlabel}%
\global\let\@eqnlabel\@vacuum}  }

\newfont{\Bbbb}{msbm7 scaled 1\@ptsize00}
\newcommand{\zs}{\raise-1pt\hbox{$\mbox{\Bbbb Z}$}}

scaled 1\@ptsize00

\font\sevenmsa=msam6 
\newfam\msafam
\textfont\msafam=\sevenmsa
\def\hexnumber@#1{\ifnum#1<10 \number#1\else
\ifnum#1=10 A\else\ifnum#1=11 B\else\ifnum#1=12 C\else \ifnum#1=13
D\else\ifnum#1=14 E\else\ifnum#1=15 F\fi\fi\fi\fi\fi\fi\fi}
\def\msa@{\hexnumber@\msafam}
\def\llcorner{\delimiter"4\msa@78\msa@78 }
\def\lrcorner{\delimiter"5\msa@79\msa@79 }
\mathchardef\blacktriangleright="3\msa@49
\mathchardef\blacktriangleleft="3\msa@4A \font\tenmsb=msbm10 scaled
1\@ptsize00
\newfam\msbfam
\textfont\msbfam=\tenmsb \scriptfont\msbfam=\tenmsb

\newdimen\Squaresize \Squaresize=14pt
\newdimen\Thickness \Thickness=0.5pt

\def\Square#1{\hbox{\vrule width \Thickness
   \vbox to \Squaresize{\hrule height \Thickness\vss
      \hbox to \Squaresize{\hss#1\hss}
   \vss\hrule height\Thickness}
\unskip\vrule width \Thickness} \kern-\Thickness}

\def\Vsquare#1{\vbox{\Square{$#1$}}\kern-\Thickness}

\def\numberbysection{\@addtoreset{equation}{section}
        \def\theequation{\thesection.\arabic{equation}}}
\numberbysection

\renewcommand{\theequation}{\thesection.\arabic{equation}}
\def\titlepage{\@restonecolfalse\if@twocolumn\@restonecoltrue\onecolumn
     \else \newpage \fi \thispagestyle{empty}\c@page\z@
        \def\thefootnote{\fnsymbol{footnote}} }

\def\endtitlepage{\if@restonecol\twocolumn \else  \fi
        \def\thefootnote{\arabic{footnote}}
        \setcounter{footnote}{0}}  
\relax

\hybrid
\makeatletter
\newdimen\normalarrayskip            
\newdimen\minarrayskip               
\normalarrayskip\baselineskip \minarrayskip\jot
\newif\ifold             \oldtrue            \def\new{\oldfalse}
\def\arraymode{\ifold\relax\else\displaystyle\fi}
\def\eqnumphantom{\phantom{(\theequation)}} 
\def\@arrayskip{\ifold\baselineskip\z@\lineskip\z@
     \else
     \baselineskip\minarrayskip\lineskip1\baselineskip\fi}

\def\@arrayclassz{\ifcase \@lastchclass \@acolampacol \or
\@ampacol \or \or \or \@addamp \or
   \@acolampacol \or \@firstampfalse \@acol \fi
\edef\@preamble{\@preamble
  \ifcase \@chnum
     \hfil$\relax\arraymode\@sharp$\hfil
     \or $\relax\arraymode\@sharp$\hfil
     \or \hfil$\relax\arraymode\@sharp$\fi}}

\def\@array[#1]#2{\setbox\@arstrutbox=\hbox{\vrule
     height\arraystretch \ht\strutbox
     depth\arraystretch \dp\strutbox
width\z@}\@mkpream{#2}\edef\@preamble{\halign \noexpand\@halignto
\bgroup \tabskip\z@ \@arstrut \@preamble \tabskip\z@ \cr}%
\let\@startpbox\@@startpbox \let\@endpbox\@@endpbox
  \if #1t\vtop \else \if#1b\vbox \else \vcenter \fi\fi
  \bgroup \let\par\relax
  \let\@sharp##\let\protect\relax
  \@arrayskip\@preamble}
\def\eqnarray{\stepcounter{equation}%
              \let\@currentlabel=\theequation
              \global\@eqnswtrue
              \global\@eqcnt\z@
              \tabskip\@centering              
              \let\\=\@eqncr
              $$%
            \halign to \displaywidth  \bgroup
             \eqnumphantom \@eqnsel
      \hskip\@centering                               
    $\displaystyle  \tabskip\z@ {##}$%
    &\global\@eqcnt\@ne \hskip 2\arraycolsep
         $ \displaystyle  \arraymode{##}$\hfil
    &\global\@eqcnt\tw@ \hskip 2\arraycolsep
         $\displaystyle\tabskip\z@{##}$\hfil
         \tabskip\@centering
    &{##}\tabskip\z@\cr}
\makeatother

\usepackage{amsthm}

\newtheorem{conj}{Conjecture}[section]

\theoremstyle{definition}
\newtheorem{ex}{Example}[section]
\newtheorem{rem}{Remark}[section]
\newtheorem{quest}{Question}[section]
\newcommand\bqa{\begin{eqnarray}}
\newcommand\eqa{\end{eqnarray}}
\def\be{\begin{eqnarray}\new\begin{array}{cc}}
\def\ee{\end{array}\end{eqnarray}}

\def\beq{\begin{equation}}
\def\eeq{\end{equation}}
\def\bse{\begin{subequations}}                
\def\ese{\end{subequations}}
\def\bp{\begin{pmatrix}}
\def\ep{\end{pmatrix}}

\def\stack#1#2{\raise0.7pt\hbox{$\mathrel{\mathop{#2}\limits^{#1}}$}}
\def\tr{\triangleright}
\def\tl{\triangleleft}
\def\sem{\mathsurround=0pt \raise1pt
\hbox{$\scriptscriptstyle>\!\!$}\:\!\!\tl}
\def\mes{\mathsurround=0pt \tr\!\:\!\raise0.8pt
\hbox{$\scriptscriptstyle\!\!<$}\,}
\def\]{\mathsurround=0pt ]\raise-2pt\hbox{$_\ast$}}

\def\<{\langle}
\def\>{\rangle}

\def\we{\raise-1pt\hbox{$\,\stackrel{\wedge}{,}\,$}}
\def\tr{{\rm tr}\,}

\newcounter{pac}[section]

\newcounter{pacc}[subsection]

0

0

\begin{document}


\vspace{15 mm}
\centerline{\Large \bf Characters, Coadjoint Orbits}
\vspace{5 mm}
\centerline{\Large \bf  and } 
\vspace{5 mm}
\centerline{\Large \bf Duistermaat-Heckman integrals}
\vspace{8 mm}
\centerline{Anton Alekseev and Samson L. Shatashvili}
\vspace{2 mm}

\vspace{5 mm}
\begin{abstract}
The asymptotics of characters $\chi_{k\lambda}(\exp(h/k))$ of irreducible representations of a compact Lie group $G$  for large values of the scaling factor $k$ are given by Duistermaat-Heckman (DH) integrals over coadjoint orbits of $G$. This phenomenon generalises to coadjoint orbits of central extensions of loop groups 
$\widehat{LG}$ and of  diffeomorphisms of the circle $\widehat{\rm Diff}(S^1)$. We show that the asymptotics of characters of integrable modules of affine Kac-Moody algebras and of the Virasoro algebra factorize into a divergent contribution of the standard form and a convergent contribution which can be interpreted as a formal DH orbital integral. 

For some Virasoro modules, our results match the formal DH integrals recently computed by Stanford and Witten.  In this case, the $k$-scaling has the same origin as the one which gives rise to   classical conformal blocks. Furthermore, we consider reduced spaces of Virasoro coadjoint orbits and we suggest a new invariant which replaces symplectic volume in the infinite dimensional situation.
We also consider other modules of the Virasoro algebra (in particular, the  modules corresponding to minimal models) and we obtain DH-type expressions which do not correspond to any Virasoro coadjoint orbits. 

We study volume functions $V(x)$ corresponding to formal DH integrals over coadjoint orbits of the Virasoro algebra. We show that they are related by the Hankel transform to spectral densities $\rho(E)$ recently studied by Saad, Shenker and Stanford.

\vspace{.5cm}


\end{abstract}

\vspace{1cm}

\newpage

\section{Introduction}

Integrals of exponential functions over coadjoint orbits can often be computed using localization formulas. One of the first examples of this phenomenon is the Harish-Chandra-Itsykson-Zuber integral \cite{HCh}, \cite{IZ} (for a review, see \cite{Tao}). For two Hermitian  $n$ by $n$ diagonal matrices $A$ and $B$ with distinct eigenvalues $a_1, \dots, a_n$  and $b_1, \dots b_n$, we have
\begin{equation}   \label{IZ}
\int_{U(n)} du \, e^{\rm Tr (A u B u^{-1})} =  
\frac{\sum_{\sigma \in S_n} (-1)^{|\sigma|} e^{\sum_i a_i b_{\sigma(i)}}}{\prod_{i<j} (a_i - a_j)(b_i-b_j)} .
\end{equation}
Here $du$ is the Haar measure on the group $U(n)$. The expression \eqref{IZ} can be understood as an integral over the coadjoint orbit $\mathcal{O}_B=\{ uBu^{-1}; u \in U(n)\}$ under the action of $U(n)$ by conjugations. Individual terms in the sum on the right hand side are in one-to-one correspondence with elements of the orbit invariant under conjugations by diagonal matrices given by permutations of eigenvalues of $B$: $B_\sigma=
{\rm diag}(b_{\sigma(1)}, \dots, b_{\sigma(n)})$.

The phenomenon described above generalises to many other cases.  In particular, it applies to orbital integrals for all compact Lie groups and also to some non-compact situations. 
One of the first papers addressing this situation is \cite{STS}. Integrals of type \eqref{IZ} are called Duistermaat-Heckman (DH) integrals \cite{DH}. More generally, this is a manifestation of localization in equivariant cohomology \cite{BV}, \cite{AB}.
There is a considerable interest in generalising this localization phenomenon to orbital integrals of infinite dimensional groups. In particular, physics motivated examples of central extensions of loop group $\widehat{LG}$ and of the diffeomorphism group of the circle $\widehat{\rm Diff}(S^1)$ attracted a lot of attention. For groups $\widehat{LG}$, there is an extensive literature both on the mathematical and on the physical sides. In \cite{Frenkel}, Frenkel suggested a Wiener type integral for $\widehat{LG}$ coadjoint orbits. 
A very similar construction appeared in \cite{Picken} in the physics literature. Recently, the probabilistic approach to $\widehat{LG}$ orbital integrals has been further developed in \cite{De1}, \cite{De2} by Bougerol and Defousseux.
The symplectic geometry approach to formal DH integral on certain $\widehat{LG}$ coadjoint orbits has been addressed by Jeffrey and Mracek \cite{Jeffrey}. 

Coadjoint of $\widehat{\rm Diff}(S^1)$ attracted attention already in the 1980s both in the framework of Representation Theory \cite{KY} and in the framework of 2-dimensional Conformal Field Theory (CFT), see \cite{W}, \cite{AS1}. Orbital integrals of the type
\begin{equation}      \label{SW}
I(\tau) = \int_{{\rm Diff}(S^1)} \mathcal{D}f \, \exp\left( \tau \int_{S^1} S(f(s)) ds\right)
\end{equation}
were recently studied by Stanford and Witten \cite{SW}. In equation \eqref{SW}, $\mathcal{D}f$ stands for some measure on the coadjoint orbit of $\widehat{\rm Diff}(S^1)$ and $S(f)$ is the Schwarzian derivative of the diffeomorphism $f$. In \cite{SW}, the integral \eqref{SW} is computed by formally applying the localization formula to the right hand side. We will denote the corresponding answer by $I^{\rm DH}(\tau)$. 

In this paper, we suggest a new approach to infinite dimensional  orbital integrals. It is based on the following observation. Let $G$ be a compact Lie group, and consider a character $\chi_\lambda(g)$ of an irreducible representation of highest weight $\lambda$. By the geometric quantisation method (or the Borel-Weil-Bott Theorem), there is a coadjoint orbit $\mathcal{O}_\lambda$ associated to this representation. We denote its dimension by ${\rm dim}(\mathcal{O}_\lambda)=2d$. Then, the character admits the following interesting asymptotic expansion:
\begin{equation}         \label{asymp_finite}
\chi_{k \lambda}\left(e^{\frac{h}{k}}\right) = \left( \frac{k}{2\pi} \right)^d \, I^{\rm DH}_{\mathcal{O}_\lambda}(h) + \dots
\end{equation}
Here $k \in \mathbb{Z}$ is a large parameter, $h \in \mathfrak{t}={\rm Lie}(T)$ is in the Lie algebra of the maximal torus $T \subset G$ and $\dots$ stands for subleading terms in $k$. Note that in \eqref{asymp_finite} we scale the highest weight $\lambda$ with a large factor $k$ and that at the same time we scale down the group element $\exp(h/k)$ so as it gets closer and closer to the group unit.

Equation \eqref{asymp_finite} can also be used to compute the value of the DH integral:
\begin{equation}
\label{revers}
I^{\rm DH}_{\mathcal{O}_\lambda}(h) = \lim_{k \to \infty} \left( \frac{2\pi}{k} \right)^d \, \chi_{k \lambda}\left(e^{\frac{h}{k}}\right).
\end{equation}
We will apply this logic to infinite dimensional orbital integrals. Our main result is roughly as follows. For each integrable representation of $\widehat{LG}$ or $\widehat{\rm Diff}(S^1)$ there are two real numbers $d, \alpha \in \mathbb{R}$ such that the asymptotic behaviour of the character is given by
\begin{equation}       \label{asymp_infinite}
\chi_{k \lambda}\left(e^{\frac{h}{k}}\right) \sim_{k \to \infty} \left( \frac{k}{2\pi} \right)^d \exp\left(\frac{ i \pi k}{12 \tau} \alpha\right) \cdot I_{\mathcal{O}_\lambda}^{\rm DH}(h).
\end{equation}
Here the parameter $\tau$ corresponds to loop rotations and we include it as one of the components of $h$.
The expression  $I_{\mathcal{O}_\lambda}^{\rm DH}(h)$  is the formal DH integral over the coadjoint orbit $\mathcal{O}_\lambda$ computed using the $\zeta$-regularisation of infinite dimensional Pfaffians. The divergent part on the right hand side includes the factor $(k/2\pi)^d$ which is similar to the 
finite dimensional situation. In examples, the ``dimension'' $d$ is not necessarily an integer, and it may be negative. There is also another exponent $\alpha$ which is the coefficient in front of the essential singularity 
$\exp(i \pi k \alpha/12\tau)$ which involves the parameter $\tau$.
In all examples that we computed, $\alpha$ is a positive integer which is equal to the number of free fields in the corresponding CFT. Asymptotic expansions \eqref{asymp_infinite} of Virasoro characters play a key role in establishing the classification of minimal models   (see pp. 357-359 in \cite{CFTbook} ).

In the case of all integrable representations of $\widehat{LG}$, one can associate a coadjoint orbit to the representation in such a way that \eqref{asymp_infinite} holds. This is a manifestation of the orbit method for $\widehat{LG}$ as established in \cite{Frenkel}. For some representations of the Virasoro algebra, equation \eqref{asymp_infinite} holds with the convergent part given by Stanford-Witten formal DH integrals over coadjoint orbits of $\widehat{\rm Diff}(S^1)$. Surprisingly, for other Virasoro representations (including those used in minimal models) we obtain a formula of type \eqref{asymp_infinite} where the convergent expression does not correspond to any coadjoint orbit. Conjecturally, these expressions are formal DH integrals over some other symplectic spaces which carry a natural action of ${\rm Diff}(S^1)$. 

In the finite dimensional situation, the DH integral serves as a generating function of volumes of reduced spaces. In more detail, the volume function
\begin{equation}
\label{fnt}
V(x) = \int_{\mathfrak{t}^*} I^{\rm DH}_{O_\lambda}(h) e^{- \langle x, h \rangle} dh
\end{equation}
is proportional to the volume of the reduced space $M_x=O_\lambda/\!\!/_x T$. In particular, $V(x) >0$ if $M_x$ is nonempty. 
We consider Fourier-Laplace transforms of formal DH integrals $I^{\rm DH}_{O_\lambda}(h)$ for coadjoint orbits of the Virasoro algebra. It turns
out that the corresponding volume functions $V(x)$ are not always positive and that they diverge at some unexpected values of $x$ ({\em e.g.} when the reduced space is a point).

To remedy these problems, we consider the Fourier-Laplace transform $V(k, x)$ of the leading term in the asymptotic expansion of the character \eqref{asymp_infinite}. It depends on $k$ and it includes the contribution which diverges for $k$ large. However, in examples that we computed, $V(k,x)$ is positive, as expected.  Furthermore, for $k$ large, $\log(V(k,x))$ has an asymptotic expansion of the form
\begin{equation}
\label{vlm}
\log(V(k,x)) = u(x) k^{\frac{1}{2}} + v(x) \ln(k) + \dots,
\end{equation}
where $\dots$ stand for subleading terms in $k$. The exponents $u(x)$ and $v(x)$ (and possibly further terms in the expansion) can be interpreted as new invariants of reduced spaces $M_x$
replacing symplectic volume. Their precise geometric meaning remains unknown.

We compare our findings with the results of \cite{SSS}. Part of their approach consists in the study of the spectral representation of formal DH integrals. In more detail, for the Virasoro algebra one has
\begin{equation}
I^{\rm DH}(\beta^{-1}) = \int_0^\infty \rho(E) e^{-\beta E} dE,
\end{equation}
where $\beta = (-i\tau)^{-1}$ and $\rho(E)$ is the density function of a quantum system (which mimics the behaviour of the formal DH integral). We observe that the volume function $V(x)$ and the spectral density $\rho(E)$ are related by a (version of) Hankel transform:
\begin{equation}
V(x) =  \, \int_0^\infty \sqrt{ \frac{E}{x}} \, I_{1}(2\sqrt{xE}) \rho(E) dE,
\end{equation}
where $I(z)$ is the modified Bessel function. Hence, $V(x)$ and $\rho(E)$ contain exactly the same amount of information. 

We would like to end the introduction with several pieces of motivation coming from the physics literature.

The relation between characters and DH integrals naturally arises in the context of path integrals. Indeed, a path integral is designed to compute a partition function of a quantum system. In the simple case when a quantum system corresponds to an irreducible representation of a (finite or infinite dimensional) group, the partition function is given by the character of the group representation on the Hilbert space of the quantum system. The quasi-classical limit corresponds to localization of the path integral on the space of classical trajectories. For compact groups, it naturally reduces to a finite dimensional integral (for examples, see \cite{AFS}). Asymptotic formulas described above are a realisation of this general principle. In the case of coadjoint orbits of $\widehat{LG}$, the corresponding quantum theory is the chiral Wess-Zumino-Witten model. In the case of coadjoint orbits of $\widehat{\rm Diff}(S^1)$, it is the gravitational Wess-Zumino action (for details, see \cite{AS1}, \cite{AS2}).

The formal DH integral \eqref{SW} arises in the Schwarzian theory motivated by the study of the SYK model \cite{EMV}, \cite{Je1}, \cite{MSY}. In this context, the scaling behaviour \eqref{asymp_infinite} appeared in \cite{Verlinde} (see also \cite{MT}). Recently, it has been discussed in many instances related to SYK, and in particular in the problem of nearly thermolized states (see {\em e.g.} \,\cite{Jensen}). It would be interesting to know whether other DH-type Virasoro integrals which do not admit  coadjoint orbit interpretation are relevant within the SYK framework.
 
 For the Virasoro algebra, the scaling limit \eqref{asymp_infinite} is related to the theory of classical conformal blocks (see \cite{ZZ}, and also \cite{zograf} for classical correlation functions). It is customary to use the Liouville theory parametrisation for the central charge:
\begin{equation}
c= 1 + 6(b+b^{-1})^2,
\end{equation}
There are two quasi-classical limits:  $b \to 0$ and $b \to \infty$. We choose to work with the first one. Then, the scaling of conformal dimensions is as follows (see \cite{ZZ}):
\begin{equation}
\Delta(\delta) \sim_{b \to 0} b^{-2} \delta,
\end{equation}
where $\delta$ is a finite constant. Note that  components of the highest weight are $(c, \Delta)$ and that they scale exactly in the same way. By putting $k \sim b^{-2}$ we obtain the natural scaling from the representation theory. 

Furthermore, it is important to mention that the classical limit of conformal blocks is intimately related to supersymmetric vacua of $N=2$ gauge theories in four space-time dimensions in the $\Omega$-background.  In the framework of AGT correspondence \cite{AGT}, $b$ is one of the parameters characterizing the $\Omega$-background.  The limit $b \to 0$ leads to the Bethe-gauge correspondence, where  a classical conformal block coincides with the twisted superpotential of the effective 2d theory as well as with the Yang-Yang functional of the corresponding Seiberg-Witten integrable system \cite{NS}.  Here we collect some references for a recent discussion of classical conformal blocks in this context:  \cite{NRS}, \cite{Piatek}, \cite{Piatek2}, \cite{Teschner1}, \cite{Teschner2}, \cite{Gorsky}.

The paper is organized as follows: in Section \ref{sec:finite_dimension}, we recall some facts about finite dimensional Duistermaat-Heckman integrals over coadjoint orbits and their relation to characters of irreducible representations. In Section \ref{sec:LG}, we briefly describe the structure of coadjoint orbits of loop groups and establish the relation between the asymptotic behaviour of characters of integrable representations and formal DH orbital integrals. In Section \ref{sec:Virasoro}, we consider coadjoint orbits of the Virasoro algebra and discuss the relation between Stanford-Witten integrals to Virasoro characters. We explain that some characters do not correspond to coadjoint orbits even though they have a DH-type asymptotic behaviour. We also discuss the positivity issue for Virasoro volume functions. In Section \ref{sec:final} we collect some open questions about infinite dimensional orbital integrals.

{\bf Acknowledgements.} Our research was triggered by discussions with E. Witten.  We are indepted to E. Meinrenken for fruitful conversations and for sharing with us his higher rank computations. We would like to thank J.-M. Bismut, P. Etingof, G. Felder, E. Getzler, K. Jensen, L. Takhtajan, F. Valach, E. Verlinde and D. Youmans for valuable discussions. The research of A.A. was supported in part by the grants number 178794, 178828 of the Swiss National Science Foundation and by the National Center of Competence in Research (NCCR) SwissMAP. The research of S.Sh. was supported by Simons Foundation under the program "Targeted Grants to Institutes". The hospitality of the CRM Montreal and of the IHES where part of this work has been done is greatly appreciated.

\section{Equivariant localization: the finite dimensional case} \label{sec:finite_dimension}

\subsection{From Geometry to Representation Theory}

In this section, we recall how localization formulas for geometric quantization of coadjoint orbits give rise to the Weyl character formula for irreducible representations of compact Lie groups. It is convenient to put this discussion in a more general context of geometric quantization of compact K\"ahler manifolds which carry an action of a compact torus.

Let $M$ be a compact K\"ahler manifold of real dimension ${\rm dim}_{\mathbb{R}}(M)=2d$ with symplectic form $\omega$. We assume that it carries a Hamiltonian action of a torus $T$ with isolated fixed points $p_1, \dots, p_m$. We denote the Lie algebra of $T$ by $\mathfrak{t}={\rm Lie}(T)$.  Weights of the $T$-action on  tangent spaces $T_{p_i}M$ at the fixed points are denoted by $w_{a,i} \in \mathfrak{t}^*$ for $a=1, \dots, d, i=1, \dots, m$. By assumption, fixed points are isolated and all the weights are non-vanishing.
Furthermore, we assume that $M$ carries a pre-quantum line bundle $L$ and that the $T$-action lifts to $L$. We denote the corresponding moment map by $\mu: M \to \mathfrak{t}^*$.

Recall the following localization formula for the $T$-equivariant index of a symplectic Dirac operator twisted by the line bundle $L$:
\begin{equation} \label{eq:index}
{\rm ind}_T(M, L)(\exp(\xi)) = 
\sum_{i=1}^m \, \frac{ e^{ i \langle \mu(p_i), \xi \rangle} }{\prod_{a=1}^n (1 - e^{- i \langle w_{a, i}, \xi \rangle}) } ,
\end{equation}
where $\xi \in \mathfrak{t}$ and $\exp(\xi) \in T$. 

Taking a tensor power $L^k$ of $L$
corresponds to scaling of the symplectic form and of the moment map:
\begin{equation}
\label{scale}
\omega \mapsto k\omega, \hskip 0.3cm \mu \mapsto k \mu.
\end{equation}
Hence, the rescaled localization formula reads,
\begin{equation}
\label{rscl}
{\rm ind}_T(M, L^k)(\exp(\xi)) = 
\sum_{i=1}^m \, \frac{ e^{ i k \langle \mu(p_i), \xi \rangle} }{\prod_{a=1}^n (1 - e^{- i \langle w_{a, i}, \xi \rangle}) } .
\end{equation}

\begin{rem}
The right hand side of equation \eqref{rscl} allows for unique reconstruction of the following geometric data: the number of fixed points, the values $\mu(p_i)$ of the moment map at  fixed points, and the list of weights $w_{a,i}$ at each fixed point. 
\end{rem}

\begin{ex}
The simplest example of the situation described above is the sphere $S^2_r$ of radius $r$ equipped with the symplectic  area form $\omega=dz \wedge d \phi$ (here $z$ and $\phi$ are cylindrical coordinates on $\mathbb{R}^3$). The manifold $S^2_r$ is pre-quantisable if $2r$ is an integer. Fixed points of the action of $T=S^1$ (by rotations around the $z$-axis) are the North pole and the South pole. The corresponding weights $w$ are equal to $+1$ and $-1$, respectively. The rescaled index formula reads
\begin{equation}  \label{eq:S^2}
{\rm ind}_{S^1}(S_r^2, L^k)(\exp(i\theta)) = \frac{e^{i k r \theta}}{1 - e^{- i\theta}} +
\frac{e^{-i k r\theta}}{1-e^{i \theta}} =\frac{ \sin\left( \left(kr+\frac{1}{2}\right))\theta\right) }{\sin\left(\frac{\theta}{2}\right)} .
\end{equation}
For $k=1$, the right hand side is the character of the irreducible representation of $G=SU(2)$ of dimension $2r+1$.
\end{ex}

More generally, consider a coadjoint orbit of a compact connected Lie group $G$ passing through a regular weight $\lambda \in \mathfrak{t}^*$. Fixed points of the maximal torus are parametrised by the elements $w \in W$ of the Weyl group of $G$. The values of the moment map at  fixed points are given by $w(\lambda)$ and the corresponding weights  are the $w$-images of positive roots $w(\alpha)$. These observations lead to the following expression for the index:
\begin{equation}
\label{ind}
{\rm ind}_T(O_\lambda, L^k)(\exp(\xi)) = \sum_{w \in W} \, \frac{e^{i k\langle w(\lambda), \xi \rangle}}{\prod_{\alpha \in \Delta_+}
(1 - e^{-i \langle w(\alpha), \xi \rangle})} .
\end{equation}
For $k=1$, this formula is a way of writing the Weyl character formula for the character of the irreducible representation of $G$ of highest weight $\lambda$.

\subsection{From Representation Theory to Geometry}

In this Section, we recall the relation between expressions for equivariant index and Duistermaat-Heckman integrals. In particular, these considerations apply to coadjoint orbits where we obtain localization formulas for orbital integrals.

Recall that on a compact symplectic manifold with a Hamiltonian $T$-action and moment map $\mu: M \to \mathfrak{t}^*$, the Duistermaat-Heckman formula computes the following oscillating integral
\begin{equation}
\label{osc}
I_M(\xi) = \int_M \frac{\omega^d}{d!} \, e^{i \langle \mu, \xi \rangle},
\end{equation}
where $\xi \in \mathfrak{t}$.
The result is given by the localization formula:
\begin{equation} \label{eq:DH}
I_M(\xi) =\left( \frac{2\pi}{\sqrt{-1}} \right)^d \sum_{i=1}^m \, \frac{e^{i \langle \mu(p_i), \xi \rangle} }{\prod_{a=1}^d \langle w_{a,i}, \xi\rangle } .
\end{equation}

\begin{rem}
The denominator of the localization formula admits the following useful interpretation. The fundamental vector field $\xi_M$  induces a linear operator $\xi_i$ on the tangent space $T_{p_i}M$ at the fixed point $p_i$. This operator is skew-symmetric with respect to the scalar product induced by the K\"ahler metric. The eigenvalues of $\xi_i$ are exactly $\langle w_{a,i}, \xi\rangle$, and we have
\begin{equation}
\label{pfn}
\prod_{a=1}^d \langle w_{a,i}, \xi \rangle = {\rm Pf}(\xi_i),
\end{equation}
where ${\rm Pf}(\xi_i)$ is the Pfaffian of the operator $\xi_i$ with respect to the natural orientation of $T_{p_i}M$ induced by the complex structure on $M$.
\end{rem}

It is well known that the Duistermaat-Heckman formula \eqref{eq:DH} can be obtained from the index formula \eqref{eq:index} by the following scaling limit:
\begin{equation}
\label{dhlim}
I_M(\xi) = \lim_{k \to \infty} \, \left( \frac{2\pi}{k} \right)^d \, {\rm ind}_T(M, L^k)\left(e^{\frac{\xi}{k}}\right) .
\end{equation}

Another way to express it is as follows:
\begin{equation}
\label{anoth}
{\rm ind}_T(M, L^k)\left(e^{\frac{\xi}{k}}\right) \sim_{k\to \infty} \left( \frac{k}{2\pi}\right)^d  I_M(\xi) + \dots .
\end{equation}
That is, after rescaling $\xi \mapsto k^{-1}\xi$, the leading $k \to \infty$ behavior of the index is described by the Duistermaat-Heckman integral.

\begin{ex}
In the example of the sphere $S^2_r$, we obtain
\begin{equation}
\label{stwo}
I_{S^2}(\theta) = \lim_{k \to \infty} \frac{2 \pi}{k} \, \frac{ \sin\left(\left(kr+\frac{1}{2}\right)\frac{\theta}{k}\right) }{ \sin\left(\frac{\theta}{2k}\right) } =
2\pi \, \frac{ e^{ir\theta} - e^{-ir\theta} }{i\theta} = \int_{S^2_r} e^{i z \theta} dz\wedge d \phi 
\end{equation}
which is exactly the DH integral for the sphere $S^2_r$ expressed in cylindrical coordinates.
\end{ex}

Similarly, for a coadjoint orbit $O_\lambda$ of a compact connected Lie group $G$ we obtain
\begin{equation}
\label{gener}
\lim_{k \to \infty} \left( \frac{2\pi}{k}\right)^d \, \sum_{w \in W} \, \frac{e^{i k\langle w(\lambda), \xi /k \rangle}}{\prod_{\alpha \in \Delta_+}
(1 - e^{-i \langle w(\alpha), \xi/k \rangle})} =
\left( \frac{2\pi}{\sqrt{-1}} \right)^d \sum_{w\in W} \frac{e^{i\langle w(\lambda), \xi \rangle}}{ \prod_{\alpha \in \Delta_+}\langle w(\alpha), \xi \rangle} .
\end{equation}
The right hand side of this equation is the Duistermaat-Heckman localization formula for the orbital integral.

\subsection{Volumes of reduced spaces}

In this Section, we recall the interpretation of Duistermaat-Heckman integrals in terms of volumes of reduced spaces. For $x \in \mathfrak{t}^*$ a regular value of the moment map $\mu: M \to \mathfrak{t}^*$, the space
\begin{equation}
M_x=\mu^{-1}(x)/T
\end{equation}
is the reduced space at level $x$. It carries a canonical symplectic form $\omega_x$. For simplicity, we assume that the torus $T$ acts freely on the level set $\mu^{-1}(x)$. In this case, $(M_x, \omega_x)$ is a symplectic manifold. 

There is the following interesting relation between the Duistermaat-Heckman integral $I_M(\xi)$ and the volume function $V(x)={\rm Vol}(M_x, \omega_x)$:
\begin{equation} \label{volume}
I_M(\xi)= {\rm Vol}(T) \int_{\mathfrak{t}^*} V(x) e^{ i \langle x, \xi \rangle} dx,
\end{equation}
where  ${\rm Vol}(T)$ is the volume of the torus $T$ with respect to the Haar measure dual to the Lebesgue measure $dx$ used in the Fourier transform. Equation \eqref{volume} implies that the volume function $V(x)$ can be recovered from the Duistermaat-Heckman integral by the inverse Fourier transform.

\begin{ex}
For $M=S^2_r$, reduced spaces for $x \in [-r, r]$ are points, and they are empty for values of $x$ outside that interval. Hence, the volume function $V(x)=\chi_{[-r,r]}(x)$ is the characteristic function of the interval $[-r,r]$.  The right hand side of equation \eqref{volume} reads
\begin{equation}
\label{nvlm}
2\pi \int_\mathbb{R} \chi_{[-r, r]}(x) e^{ix \theta} dx = 2\pi \int_{-r}^{r} e^{ix\theta} dx = 2\pi \, \frac{e^{ir\theta} - e^{-ir\theta}}{i \theta} = I_{S^2_r}(\theta),
\end{equation}
as required. For the inverse Fourier transform, we compute
\begin{equation}
V(x) = \frac{1}{2\pi} \int_{\mathbb{R} + i \epsilon} \frac{e^{i(r-x) \theta} - e^{-i(r+x)\theta}}{i\theta} \, d\theta=
\left\{
\begin{array}{ll}
1 & {\rm if} \,\, -r < x < r \\
0 & {\rm otherwise}
\end{array}
\right.
\end{equation}
Note that the integration contour is shifted to avoid the pole at $\theta=0$. For $x < -r$, the contour can be contracted in the upper half plane and the integral vanishes. For $x >r$, both exponentials contribute in the residue at $\theta=0$ and their contributions cancel each other. For $-r < x < r$, only the second exponential contributes which yields $V(x)=1$.
\end{ex}

\begin{rem}
In general, the Duistermaat-Heckman Theorem implies that $V(x)$ is a piece-wise polynomial function. Fixed points and other submanifolds with nontrivial stabilisers determine its wall-crossing structure (for details, see \cite{GLS}). For instance, in the example of a sphere $S^2_r$ the function $V(x)$ jumps at $x=\pm r$, the images of the fixed points under the moment map.
\end{rem}

\subsection{Scaling and Representation Theory}

We end  this Section with a brief discussion of scaling from the view point of Representation Theory. Let $\{ t^a \}$ be a basis of the Lie algebra $\mathfrak{g}$. The Lie brackets are of the form
\begin{equation}
\label{alg}
[t^a, t^b] = f^{ab}_c t^c,
\end{equation}
where $f^{ab}_c$ are  structure constants of $\mathfrak{g}$. Consider the scaling transformation $t^a=k \tau^a$. Lie brackets of the rescaled generators $\tau^a$ acquire the form
\begin{equation}
\label{scalealg}
[\tau^a, \tau^b] =k^{-1}\,  f^{ab}_c \tau^c.
\end{equation}
With the usual interpretation $\hbar=1/k$, we obtain $[\tau^a, \tau^b]=\hbar f^{ab}_c \tau^c$. In the $\hbar \to 0$ limit, equation \eqref{scalealg} gives rise to Poisson brackets
\begin{equation}
\label{pois}
\{ \tau^a, \tau^b\} =\lim_{\hbar \to 0} \frac{[\tau^a, \tau^b]}{\hbar} = f^{ab}_c \tau^c.
\end{equation}
Consider an example of $G=SU(2)$. In the representation of spin $J$, the quadratic Casimir operator $C=t^at^a$ has an eigenvalue $C_J=J(J+1)$. Under the scaling transformation, we get $C=k^2 \tau^a \tau^a=k^2c$. If we scale the highest weight as $J=kj$ (similar to the prescription of the previous Sections), we obtain the following scaling of eigenvalues of the operator  $c = k^{-2} C$:
\begin{equation}
\label{casm}
\frac{C_{kj}}{k^2}=\frac{ (kj)(kj+1)}{k^2} \to_{k \to \infty} j^2.
\end{equation}
This is a general rule: when passing to the scaling limit, all weights and highest weights scale as $J=kj$.

\section{Coadjoint orbits of loop groups} \label{sec:LG}

In the finite dimensional case, geometric quantization provides a link between Geometry and Representation Theory. In the previous Section, we recalled that  the equivariant index of a coadjoint orbit coincides with the character of the corresponding irreducible representation. This fact is a manifestation of the Borel-Weil-Bott Theorem which states that irreducible representations are realized in spaces of holomorphic sections of pre-quantum line bundles over coadjoint orbits. Furthermore, the scaling limit of the index gives rise to the oscillating Duistermaat-Heckman integral.

In the infinite dimensional case, most of the notions mentioned above no longer exist (or they are difficult to define). It applies to all geometric objects such as indices (of Dirac operators) and Duistermaat-Heckman integrals which would involve defining a Liouville measure on infinite-dimensional spaces. At the same time, in many cases characters of irreducible representations are well defined and well behaved objects. In what follows, our strategy is to use them to gain insight into geometric properties of coadjoint orbits (and possibly of other spaces).

\subsection{Geometry of coadjoint orbits of $\widehat{LG}$}

As before, let $G$ be a compact connected Lie group. Furthermore, assume that $G$ is simply connected. Then, the loop group $LG$ is defined as the connected space of maps from the circle $S^1$ to $G$ equipped with the pointwise multiplication. Given an invariant scalar product on the Lie algebra $\mathfrak{g}={\rm Lie}(G)$, the loop group has a canonical central extension $\widehat{LG}$. The dual of the corresponding Lie algebra is modelled by pairs $(A, l)$, where $A \in \Omega^1(S^1, \mathfrak{g})$ and $l \in \mathbb{R}$.
The coadjoint action of $LG$ is by gauge transformations. In more detail, for $g(s) \in LG$ and $A=a(s) ds$ the action reads
\begin{equation}
\label{gg}
g: a \mapsto gag^{-1} + l(\partial_sgg^{-1}), \hskip 0.3cm l \mapsto l.
\end{equation}
Coadjoint orbits of $\widehat{LG}$ are parametrised by conjugacy classes $\mathcal{C} \subset G$:
\begin{equation}
\label{cjn}
\mathcal{O}_\mathcal{C} = \{ (A, l); {\rm Hol}(S^1, l^{-1}A) \in \mathcal{C}\} .
\end{equation}
Here we view $l^{-1} A$ as a connection on the trivial $G$-bundle over the circle, and ${\rm Hol}(S^1, l^{-1}A)$ stands for its holonomy. 

Let $T \subset G$ be a maximal torus and $\mathfrak{t} = {\rm Lie}(T)$ its Lie algebra. For each conjugacy class $\mathcal{C} \subset G$, there is a unique element $u_\mathcal{C} \in \mathfrak{a} \subset \mathfrak{t}$ in the Weyl alcove such that $\exp(u_\mathcal{C}) \in \mathcal{C}$. Furthermore, all the other elements of $u \in \mathfrak{t}$ such that $\exp(u) \in \mathcal{C}$ form an orbit of $u_\mathcal{C}$ under the action of the affine Weyl group:
\begin{equation}
\label{shft}
u = w(u_\mathcal{C}) + v,
\end{equation}
where $w \in W$ is an element of the  Weyl group of $G$ and $v \in \Lambda$ is an element of the lattice $\Lambda = {\rm ker}(\exp: \mathfrak{t} \to T)$.

Every coadjoint orbit $\mathcal{O}_\mathcal{C}$ carries an action of the torus $T \times S^1$:
\begin{equation}   \label{action_KMorbits}
(t, \theta): (a, l) \mapsto (ta(s+ \theta)t^{-1}, l).
\end{equation}
This action is Hamiltonian with respect to the canonical symplectic structure on $\mathcal{O}_\mathcal{C}$. The moment map is of the form
\begin{equation}
\label{mmp}
\mu(a, l)=(\mu_T, \mu_{S^1})= \left(  \pi_\mathfrak{t}\left(\int_{S^1} a(s) ds \right) , \frac{1}{2l} \int_{S^1} {\rm Tr}(a(s)^2) ds \right) .
\end{equation}
Here the first component of $\mu$ takes values in $\mathfrak{t}^* \cong \mathfrak{t}$, $\pi_\mathfrak{t}: \mathfrak{g} \to \mathfrak{t}$ is the orthogonal projection, ${\rm Tr}$ stands for the invariant scalar product on $\mathfrak{g}$ and the second component of the moment map is the classical counterpart of the Sugawara construction.

The fixed points of the action \eqref{action_KMorbits} are of the form
\begin{equation}
\label{fxp}
\left(A= \frac{l}{2\pi} (w(u_\mathcal{C}) + v) ds, l\right),
\end{equation}
and the corresponding values of the moment map read
\begin{equation}
\label{mmpf}
\mu_{w,v} = \left( l (w(u_\mathcal{C}) + v), \frac{l}{4\pi} {\rm Tr}(w(u_\mathcal{C}) + v)^2 \right).
\end{equation}
In what follows we will mainly focus on the example of $G=SU(2)$ that we consider in more detail:

\begin{ex}
For $G=SU(2)$, the Weyl alcove $\mathfrak{a} \cong [0,\pi]$ can be parametrised by $0 \leq \phi \leq \pi$, and the corresponding conjugacy classes are passing through the diagonal matrices
\begin{equation}
\label{mtrc}
\left(
\begin{array}{ll}
e^{i\phi} & 0 \\
0 & e^{- i \phi}
\end{array}
\right) \in \mathcal{C}.
\end{equation}
On the  coadjoint orbit $O_\mathcal{C}$,  fixed points of the $T \times S^1$ action are of the form
\begin{equation}
\label{sfx}
\left( 
\frac{\sqrt{-1} l}{2\pi} 
\left( 
\begin{array}{cc}
\pm \phi + 2\pi n & 0 \\
0 & \mp \phi - 2 \pi n
\end{array}
\right) ds, l
\right) 
\end{equation}
for $n \in \mathbb{Z}$. Values of the moment map at  fixed points are given by
\begin{equation}
\label{mmfxp}
\mu_{\pm, n} = 
\left( 
\sqrt{-1} l 
\left( 
\begin{array}{cc}
\pm \phi + 2\pi n & 0 \\
0 & \mp \phi - 2 \pi n
\end{array}
\right), - \frac{l}{2\pi} (\pm \phi + 2 \pi n)^2 \right).
\end{equation}

\end{ex}

\subsection{Orbital integrals for $\widehat{LG}$}

In this Section, we discuss orbital integrals over coadjoint orbits of 
$\widehat{LG}$. Again, we will focus on the case of $G=SU(2)$.

\subsubsection{Formal Duistermaat-Heckman integrals}
While it is difficult to make sense of the Liouville measure in the infinite dimensional case, one can take the fixed point formula as a definition of the Duistermaat-Heckman integral. In presenting this approach we follow the ideas of \cite{Jeffrey}. The formal Duistermaat-Heckman integral over a coadjoint orbit of $\widehat{LG}$ depends on two parameters,  $b \in \mathfrak{t}$ and $t \in \mathbb{R}$ corresponding to two components of the moment map. The definition, inspired by the localization formula, reads
\begin{equation} \label{formalDH}
I^{\rm DH}(b, t) = \sum_p \frac{e^{{\rm Tr}(b \mu_T(p)) + t \mu_{S^1}}}{{\rm Pf}(A_{b, t}(p))}.
\end{equation}
Here the sum is over fixed point of the $T \times S^1$ action on the corresponding coadjoint orbit, and $A_{b, t}(p)$ is the linear operator representing the fundamental vector field $\xi_{b, t}$ on the tangent space to the fixed point $p$.

The formal expression \eqref{formalDH} has two potential problems. First, the operator $A_{b, t}(p)$ acts on an infinite dimensional space. Hence, it is not obvious how to define the Pfaffian. Second, the set of fixed points on a coadjoint orbit is infinite and the sum in \eqref{formalDH} is an infinite sum which may have convergence problems. In the case of $G=SU(2)$, we consider these two issues in detail.

First, for all $p$ the operator $A_{b, t}$ is of the form
\begin{equation}
\label{allp}
A_{b, \tau}= t \partial_s + [ b, \cdot].
\end{equation}
It acts on the space
\begin{equation}
\label{spcs}
T_p\mathcal{O}_\mathcal{C} \cong L\mathfrak{g}/\mathfrak{g}_{u_\mathcal{C}},
\end{equation}
where $\mathfrak{g}_x=\{ y \in \mathfrak{g}; [x,y]=0\}$ is the stabilizer of $p$. 
For $G=SU(2)$, $\tau \in \mathbb{R}_+$ and 
\begin{equation}
\label{sutwo}
b= \sqrt{-1}
\left(
\begin{array}{rr}
\frac{B}{2} & 0  \\
0 & -\frac{B}{2}
\end{array}
\right)
\end{equation}
with $\beta \notin \mathbb{Z}$, the eigenvalues of the operator $A_{b, \tau}(p)$ are equal to 
$\sqrt{-1}(n t), \sqrt{-1}(n t +  B), \sqrt{-1}(nt - B)$ for $n \in \mathbb{Z}_{\neq 0}$, and there are two more eigenvalues: $B$ and $-B$ which correspond to $n=0$.
The $\zeta$-regularization of the determinant of $A_{b, \tau}(p)$ can be computed as follows
\begin{equation}
\label{zreglz}
{\rm det}_\zeta(t \partial_s + [b, \cdot]) = B^2 {\rm det}'_\zeta(t \partial_s) {\rm det}'_F(1 + (t\partial_s)^{-1}[b, \cdot])
\end{equation}
where ${\rm det}'$ stands for the determinant over the subspace orthogonal to constant loops, $B^2$ is the determinant of the (dimension 2) operator acting on constant loops, and ${\rm det}'_F$ is the Fredholm determinant of the bounded operator $(1 + (t \partial_s)^{-1}[b, \cdot])$ (see Chapter 5.5 in \cite{Ta} for details). Furthermore, we have
\begin{equation}
\label{dz}
{\rm det}'_\zeta(t \partial_s)=\left(\frac{2\pi}{t}\right)^3, \hskip 0.2cm 
{\rm det}'_F(1 + (t \partial_s)^{-1}[b, \cdot])=
\prod_{n=1}^\infty \left(1 -\frac{B^2}{t^2 n^2}\right)^2 =
\left(\frac{t}{\pi B}\right)^2 \sin^2\left(\frac{\pi B}{t}\right)
\end{equation}
which yields
\begin{equation}
\label{dtz}
{\rm det}_\zeta(A_{b, t}(p)) = 
\frac{8\pi}{ t} \sin^2\left(\frac{\pi B}{t}\right) .
\end{equation}
The $\zeta$-regularized Pfaffian reads
\begin{equation}
\label{pfzf}
{\rm Pf}_\zeta(A_{b, t}(p)) = \pm \frac{2 \sqrt{2\pi}}{ t^{1/2}} \sin\left(\frac{\pi B}{t}\right),
\end{equation}
where the sign is positive for fixed points with labels $(+, n)$ and negative for fixed points with labels $(-,n)$.

\begin{rem}
In \cite{Picken} and \cite{Jeffrey}, the Pfaffian is regularized in a somewhat different way:  the operator $A_{b, t}(p)$ is replaced by a Fredholm operator (by dividing by the unbounded operator $t \partial_s$), and then one  computes  the Fredholm determinant. 
\end{rem}

Next, we address the convergence question for the infinite sum. In the case of $G=SU(2)$, it reads:
\begin{equation}
\label{convsm}
I^{\rm DH}(B, t)=
\frac{t^{1/2}}{2\sqrt{2\pi}\sin(\pi B/t)} \sum_{n\in \mathbb{Z}} \left( e^{-\frac{lt}{2\pi}(2\pi n + \phi)^2 - lB(2\pi n + \phi)} -e^{-\frac{lt}{2\pi}(2\pi n - \phi)^2 - lB(2\pi n - \phi)} \right).
\end{equation}
By changing the sign of the dummy variable $n$ in the second term, we can rewrite this expression in the following form:
\begin{equation}
\label{rwrt}
I^{\rm DH}(B, t)=
\frac{t^{1/2}}{2 \sqrt{2\pi} \sin(\pi B/t)} 
e^{-\frac{l t \phi^2}{2\pi}} 
\sum_{n\in \mathbb{Z}} e^{-2l(\pi n^2 + n \phi)t} 
\left( e^{-lB(2\pi n + \phi)} - e^{lB(2\pi n + \phi)} 
\right) .
\end{equation}
The infinite sum is convergent for ${\rm Re}(t) >0$. In particular, the right hand side is well defined for $ t > 0$. There are no restrictions on the values of $B$.
This situation is in contrast with the case of localization formulas which contain a finite number of terms: they are usually defined for all generic values of parameters.

\begin{rem}
The volume function for a certain $\widehat{LSU(2)}$ coadjoint orbit is studied in \cite{Jeffrey} using the theory of hyperfunctions. We will discuss some infinite dimensional volume functions in the next Section using coadjoint orbits of the Virasoro algebra as examples.
\end{rem}

\begin{rem}
In \cite{Frenkel}, Frenkel suggested another approach to orbital integrals for loop groups. It turns out that an orbital integral over $\mathcal{O}_\mathcal{C}$ can be interpreted as a Wiener integral over continuous paths $g_s: [0,1] \to G$ with $g_0=e$ and  $g_1 \in \mathcal{C}$. Such an integral is given by the heat kernel $p_t(g_0, g_1)$ on the group. The formula in terms of fixed points is then obtained by  Poisson resummation (or Macdonald identity for higher rank). For more recent work following the probabilistic approach, see \cite{De1}, \cite{De2}.
\end{rem}

\begin{rem}
In \cite{Bismut}, Bismut  showed that there is a 1-parameter family of operators $L_b$ on the tangent bundle $TG$ of the group $G$ such that the trace of the heat kernel of $L_b$ is independent of $b$. Furthermore, in the limit $b\to 0$ the heat kernel of $L_b$ approaches the heat kernel of the Laplacian on $G$ giving rise to the Frenkel formula. In the limit $b \to \infty$, the operator $L_b$ approaches the geodesic vector field on $TG$ which explains  localization to fixed points on the coadjoint orbit of $LG$.
\end{rem}

%
%
%
%

\subsection{Scaling of $\widehat{LG}$ characters}
We end this Section with the approach to $\widehat{LG}$ orbital integral which uses character formulas for integrable modules. Recall that affine Kac-Moody Lie algebras are defined by  the  Lie bracket
\begin{equation}
\label{kmalg}
[J^a_m, J^b_n] = f^{ab}_c J_{m+n}^c + \delta_{m+n, 0} l,
\end{equation}
where $a,b,c$ label a basis of a finite dimensional Lie algebra $\mathfrak{g}$, $f^{ab}_c$ are its structure constants, and $l$ can either be viewed as a central generator, or as a number (the level). Usually, it is convenient to add a generator $L_0$ given by the Sugawara construction which generates an $S^1$-action by rotation of a loop: $[L_0, J^a_n]=nJ^a_n$.

For $\mathfrak{g}={\rm su}(2)$ and $l$ a positive integer, the Kac-Moody algebra $\widehat{{\rm su}(2)}_l$
admits $l+1$ integrable irreducible representations labeled by spins $j=0, 1/2, \dots, l/2$. The Kac-Weyl character formula gives characters of these representations:
\begin{equation}
\label{kw}
\chi_{l,j}(\tau, \gamma) ={\rm Tr}_{V_{l,j}}\left( e^{2\pi i \tau L_0+ i \gamma J_0^z}\right) = \frac{\Delta_{l,j}(\tau, \gamma)}{D_l(\tau, \gamma)},
\end{equation}
where the  denominator does not depend on $j$:
\begin{equation}
\label{dmntr}
D_l(\tau, \gamma) = (1 - e^{-i\gamma})\prod_{r=1}^\infty(1-e^{2\pi i r \tau})(1-e^{2\pi i r \tau} e^{i \gamma})(1 - e^{2 \pi i r \tau} e^{-i\gamma}),
\end{equation}
and the numerator is given by formula
\begin{equation}
\label{num}
\Delta_{l,j}(\tau, \gamma)= e^{2\pi i \tau j(j+1)/(l+2)} \sum_{n \in \mathbb{Z}}
e^{2\pi i \tau((l+2)n^2 + n(2j+1))} ( e^{i(j+(l+2)n)\gamma} - e^{-i(j+1+(l+2)n)\gamma}).
\end{equation}
We will be interested in the following question: can one define a scaling of the representation parameters
$(l, j)$ and of the Lie algebra parameters $(\tau, \gamma)$  in such a way that the asymptotic behaviour of the character formula gives rise to a Duistermaat-Heckman type expression. Inspired by the finite dimensional case, we introduce the following scaling:
\begin{equation}
\label{kmscal}
l \mapsto kl, j\mapsto kj, \tau \mapsto \frac{\tau}{k}, \gamma \mapsto \frac{\gamma}{k}.
\end{equation}
First, we analyse the behaviour of the numerator under the scaling transformation:
\begin{equation}
\label{dscal}
\Delta_{kl,kj}\left(\frac{\tau}{k}, \frac{\gamma}{k}\right) \rightarrow_{k\to \infty} e^{2\pi i \tau j^2/l} \sum_{n \in \mathbb{Z}}
e^{2\pi i \tau((ln^2 + 2j n)} ( e^{i(j+ln)\gamma} - e^{-i(j+ln)\gamma}).
\end{equation}
Note that the numerator has a well defined limit, and that all the terms of the Kac-Weyl formula are still present with slightly simplified exponents.

Now we turn to the large $k$ behaviour of the denominator.
Recall that the Dedekind $\eta$-function $\eta(\tau)= q^{1/24} \prod_{n=1}^\infty (1-q^n)$, where $q=\exp(2\pi i \tau)$ and ${\rm Im} \, \tau >0$ verifies the modular property
\begin{equation}
\label{ded}
\eta(-1/\tau) = \sqrt{-i\tau} \eta(\tau).
\end{equation}
This implies
\begin{equation}
\label{subs}
\frac{1}{\prod_{n=1}^\infty (1 - q^n)} = \frac{1}{\prod_{n=1}^\infty (1 - e^{2\pi i n \tau})} =
\frac{\sqrt{-i\tau} \, e^{\pi i(\tau + \tau^{-1})/12}}{\prod_{n=1}^\infty (1 - e^{-2\pi i n/\tau})} .
\end{equation}

Furthermore, recall the following identity:
\begin{equation}        \label{identity}
(1 - e^{-i\gamma})\prod_{l=1}^\infty(1-e^{2\pi i l \tau} e^{i \gamma})(1 - e^{2 \pi i l \tau} e^{-i\gamma})=(q^x,q)_\infty (q^{1-x}, q)_\infty,
\end{equation}
where  $(a,q)_\infty=\prod_{k=0}^\infty(1-aq^k)$ is the Pochhammer symbol and $x=-\gamma/2\pi \tau$.
The combination of Pochhammer symbols on the right hand side of \eqref{identity} has the following asymptotic behaviour  (see Corollary 3.3 in \cite{Lambert} and Corollary 1.3 in \cite{Katsurada}):
\begin{equation}
\label{pohone}
(q^x,q)_\infty(q^{1-x},q)_\infty \sim_{q \to 1} 2 \sin(\pi x) e^{\pi^2/3\ln(q)} q^{-1/2(1/6-x+x^2)} .
\end{equation}
Applying rescaling \eqref{kmscal}, we obtain the asymptotic behaviour of the inverse denominator for $k\to \infty$: 
\begin{equation}
\label{pohtwo}
D_{kl}\left(\frac{\tau}{k}, \frac{\gamma}{k}\right)^{-1} \sim_{k\to \infty} \frac{1}{2 \sin(\gamma/2\tau)} \left( \frac{- i \tau}{k}\right)^{1/2} e^{i \pi k/4 \tau} =
\left( \frac{2\pi}{k} \right)^{1/2} e^{i\pi k/4 \tau} \cdot  \frac{(-i \tau)^{1/2}}{2 \sqrt{2\pi} \sin(\gamma/2\tau)}.
\end{equation}
The divergent factor is of the form $(k/2\pi)^d \exp(i \pi \alpha k/12 \tau)$ with $d=-1/2$ and $\alpha=3$. The factor  $(k/2\pi)^d$ is similar to the finite dimensional case even though the ``dimension'' $d$ is not an integer and negative. The factor $\exp(i \pi \alpha k/12 \tau)$ has an essential singularity at $k=\infty$, and this singularity depends on the Lie algebra parameter $\tau$.

\begin{rem}
A possible CFT interpretation of the exponent $\alpha$ is the number of free fields in the theory. Indeed, irreducible representations of $\widehat{LSU(2)}$ can be resolved in terms of free field Wakimoto modules with 3 free fields. This matches higher rank calculations of E. Meinrenken \cite{Meinrenken}.
\end{rem}

 In conclusion, the behaviour of the character under the scaling transformation is given by formula
\begin{equation}
\label{scaltwo}
\chi_{kl,kj}\left(\frac{\tau}{k}, \frac{\gamma}{k}\right) \sim_{k\to \infty} 
\left( \frac{2\pi}{k} \right)^{1/2} e^{i\pi k/4 \tau} 
 \,  \cdot  I_{l,j}(\tau, \gamma),
\end{equation}
where 
\begin{equation}
\label{finkm}
I_{l,j}(\tau, \gamma)= \frac{(-i \tau)^{1/2}}{2 \sqrt{2\pi} \sin(\gamma/2\tau)} e^{2\pi i \tau j^2/k} \sum_{n \in \mathbb{Z}}
e^{2\pi i \tau((kn^2 + 2j n)} ( e^{i(j+kn)\gamma} - e^{-i(j+kn)\gamma}).
\end{equation}
It is easy to see that after substitutions
\begin{equation}
j = \frac{l \phi}{2 \pi}, \hskip 0.2cm \tau = it, \hskip 0.3cm \gamma = 2\pi i B,
\end{equation}
we obtain the expression for the formal DH integral:
\begin{equation} \label{DH=char}
I_{l, l\phi/2\pi}(it, 2\pi i B) = I^{\rm DH}_{l, \phi}(t, B).
\end{equation}

\begin{rem}
The equation \eqref{DH=char} establishes a link between two approaches to orbital integrals: through formal Duistermaat-Heckman formulas and through asymptotic behaviour of characters. Note that  DH formulas are well-defined and unambiguous once we chose the $\zeta$-regularisation of Pfaffians. 
\end{rem}

\section{Character formulas and localization for Virasoro algebra} \label{sec:Virasoro}

In this Section, we consider coadjoint orbits of the Virasoro algebra. We use character formulas as a starting point and we derive Duistermaat-Heckman type expressions from the asymptotic expansion. In some cases, we match the results with Duistermaat-Heckman integrals of Stanford-Witten. In other cases, we discover Duistermaat-Heckman formulas which probably correspond to some new symplectic spaces. We also consider Virasoro volume functions and their surprising properties.

\subsection{Coadjoint orbits of $\widehat{\rm Diff}(S^1)$}
One can think of elements of the group ${\rm Diff}(S^1)$ as smooth maps $f: s \mapsto f(s)$ such that 
$f'(s) >0$ and $f(s+2\pi) = f(s) + 2 \pi$. The corresponding Lie algebra consists of vector fields on the circle
$L= L(s) \partial_s$ with $L(s) = \sum_{n\in \mathbb{Z}} L_n \exp(in s)$. It admits a universal central extension, the Virasoro algebra ${\rm Vir}$:
\begin{equation} \label{Vir}
[L_m, L_n] = (m-n) L_{m+n} + \frac{c}{12} (m^3-m) \delta_{m+n, 0},
\end{equation}
where $c$ is a central generator. The Virasoro Lie algebra contains the following interesting ${\rm sl}(2)$ Lie subalgebras:
\begin{equation}
\label{sltwosub}
{\rm sl}(2)_m= \mathbb{C}\left\langle E_m=\frac{L_m}{m}, F_m=\frac{L_{-m}}{m}, H_m=\frac{L_0}{m} + \frac{c(m^2-1)}{24m} \right\rangle.
\end{equation}

The central extension \eqref{Vir} gives rise to a central extension of ${\rm Diff}(S^1)$ denoted $\widehat{\rm Diff}(S^1)$. The geometry of coadjoint orbits of $\widehat{\rm Diff}(S^1)$ was pioneered in \cite{LP}, \cite{Kir}.
One can identify the dual of the Lie algebra ${\rm Vir}$ with the space of pairs $(b, c)$, where $c \in \mathbb{R}$ is the central charge and $b=b(s)ds^2$ is a quadratic differential on the circle. The coadjoint action looks as follows:
\begin{equation}
\label{quaddifl}
f: (b, c) \mapsto \left( b^f(s)ds^2= \left(b(f(s))f'(s)^2 + \frac{c}{12}S(f)\right) ds^2, c\right),
\end{equation}
where $S(f)$ is the Schwarzian derivative of $f$:
\begin{equation}
\label{schwrzn}
S(f) = \frac{f'''}{f'} - \frac{3}{2} \left( \frac{f''}{f'} \right)^2 .
\end{equation}

Coadjoint orbits are of the form
\begin{equation}
\label{soneacn}
\mathcal{O}_b= \{ b^f(s)ds^2; f \in {\rm Diff}(S^1)\}.
\end{equation}
They carry a symplectic form and an $S^1$-action 
$$
\theta: b(s)ds^2 \mapsto b(s+\theta)ds^2
$$
with moment map
\begin{equation}
\label{mmpvir}
\mu(b) = \int_{S^1} b(s) ds.
\end{equation}
Fixed point of the $S^1$-action are given by $b(s) = {\rm const}$. We will be interested in coadjoint orbits passing through such constant quadratic differentials. An orbit of $b(s)=b_0$ contains exactly one $S^1$ fixed point which coincides with $b_0$. 

Stabilisers of coadjoint orbits $\mathcal{O}_{b_0}$ depend on the constant $b_0$. For generic values of $b_0$, the stabiliser subgroup is equal to $S^1$. For special values
\begin{equation}
\label{spclbno}
b_0 = \frac{c m^2}{24},
\end{equation}
the stabiliser is isomorphic to ${\rm SL}(2, \mathbb{R})_m$ integrating the Lie subalgebra ${\rm sl}(2, \mathbb{R})_m$.

For $c>0$ and  $b_0 \leq c/24$, the moment map \eqref{mmpvir} is bounded from above, and the $S^1$ fixed point is the maximum of $\mu(b)$. For the values $b_0 > c/24$, the $S^1$ fixed point is a saddle point of 
$\mu(b)$ and the moment map is unbounded. In more detail, for $f(x)=x + \sum_{n \in \mathbb{Z}} f_n e^{in s}$ with $f_{-n} = \bar{f}_n$, we obtain
\begin{equation} \label{quadraticform}
\mu(b_0^f) =2\pi b_0 +  4\pi \sum_{n =1}^\infty \left(b_0 n^2 - \frac{c}{24} n^4\right) |f_n|^2 + \dots,
\end{equation}
where $\dots$ stand for higher order terms in $f_n$'s. For $n^2 > 24b_0/c$, the quadratic form \eqref{quadraticform} is negative definite. If $b_0 > c/24$, then for   a finite number of values of $n$ with $n^2 < 24c/b_0$ the quadratic form \eqref{quadraticform} is positive definite giving rise to a saddle point.

\subsection{Stanford-Witten orbital integrals}
In this Section, it is our aim to define and study orbital integrals for the Virasoto algebra. Formally, they look as follows:
\begin{equation}
\label{virorbint}
I^{\rm DH}_{b_0}(t) = \int_{\mathcal{O}_{b_0}} d\nu(b) e^{t \mu(b)},
\end{equation}
where $b \in \mathcal{O}_{b_0}$ is a point on the orbit, and $d\nu(b)$ is the integration measure to be determined.

Formal DH integrals for Virasoro coadjoint orbits were recently studied by Stanford-Witten \cite{SW}. For generic values of $b_0$, the $S^1$-action at the  fixed point $b(s)=b_0$ gives rise to the linear operator $A_t=t \partial_s$. Its $\zeta$-regularised Pfaffian is given by
\begin{equation}
\label{pfvir}
{\rm Pf}_\zeta(A_t) =\left( \frac{2\pi}{t} \right)^{1/2}.
\end{equation}
This gives rise to a formal DH integral of the form
\begin{equation}
\label{dvir}
I^{\rm DH}_{b_0}(t) = \frac{e^{\mu(p) t}}{ {\rm Pf}_\zeta(A_t) } = \frac{t^{1/2} }{\sqrt{2\pi} } \, e^{2\pi b_0 t},
\end{equation}
where $\mu(p) = 2\pi b_0$ is the value of the moment map at the unique fixed point. 

Note that the form of the formal DH integral does not depend on whether $b_0 < c/24$ or $b_0>c/24$, and whether the only fixed point 
is a maximum or a saddle point of the moment map. We will return to this question when we discuss Virasoro representations.

For the exceptional orbit with stabiliser ${\rm SL}(2, \mathbb{R})_m$, the operator $A^{(m)}_t$ acts on the space
\begin{equation}
\label{stblzr}
V_m =\left\{ \sum_n v_n e^{ins}; v_0=v_{\pm m}=0 \right\}.
\end{equation}
The corresponding $\zeta$-regularised Pfaffian is of the form
\begin{equation}
\label{virregpf}
{\rm Pf}_\zeta(A_t^{(m)}) = (mt)^{-1} \left( \frac{2\pi}{t} \right)^{1/2} = \frac{\sqrt{2\pi}}{m t^{3/2}}.
\end{equation}
and the formal DH integral is given by
\begin{equation}
\label{frmlans}
I^{\rm DH}_m(t) = \frac{ e^{\mu^{(m)}(p) t} }{ {\rm Pf}_\zeta(A_t^{(m)}) } = \frac{m t^{3/2}}{ \sqrt{2\pi}} \, 
e^{ \frac{\pi m^2 ct}{12}} \,  .
\end{equation}
Again, the form of the answer is independent of whether $m=1$ and the moment map is bounded from above or $m>1$ and the moment map is unbounded.

Note the following interesting relation between formal DH integrals of different types:
\begin{equation}        \label{2DH}
I^{\rm DH}_m(t)=\frac{m}{2\pi} \, \left. \frac{\partial I^{\rm DH}_{b_0}(t)}{\partial b_0}\right|_{b_0=\frac{m^2c}{24}}
\end{equation}
For $m=1$, it can be interpreted as a relation between partition functions of the Jackiw-Teitelboim gravity model on the annulus (the integral $I^{\rm DH}_{b_0}(t)$) and on the disk (the integral $I_1^{\rm DH}(t)$), see \cite{SSS}.

\subsection{Asymptotic behaviour of Virasoro characters}
In this Section, we consider asymptotic behaviour of characters 
for different types of modules of the Virasoro algebra.
In some cases, it matches Stanford-Witten orbital integrals of the previous Section. In other cases, one obtains the DH type behaviour which does not match any coadjoint orbit of the Virasoro algebra.

\subsubsection{Irreducible Verma modules}
First, consider a Virasoro Verma module $V_{(c, \Delta)}$, and assume that it is irreducible. Then, the corresponding character is given by formula
\begin{equation}
\label{vrichrctr}
\chi_{(c, \Delta)}(\tau) = \frac{q^{\Delta-c/24}}{\prod_{n=1}^\infty (1 - q^n)},
\end{equation}
where $q=\exp(2\pi i \tau)$, and ${\rm Im} \, \tau >0$.

%
%
%
The scaling transformation reads:
\begin{equation}
\label{orscl}
\Delta \mapsto k \Delta, \hskip 0.3cm c \mapsto k c, \hskip 0.3cm \tau \mapsto \frac{\tau}{k}.
\end{equation}
For the character, this rescaling yields the following large $k$ asymptotic formula:
\begin{equation}
\label{lrgscl}
\chi_{(kc, k\Delta)}\left(\frac{\tau}{k}\right) \sim_{k \to \infty} \frac{\sqrt{2\pi} e^{\frac{\pi i k}{12 \tau}}}{\sqrt{k}} \,  \cdot  \frac{\sqrt{-i \tau}}{\sqrt{2\pi}}  \, e^{2\pi i (\Delta - c/24) \tau}  = \left(\frac{2\pi}{k}\right)^{1/2} e^{\frac{\pi i k}{12 \tau}}  \cdot I_{(c, \Delta)}(\tau).
\end{equation}
Here the divergent part has exponents $d=-1/2$ and $\alpha=1$ (see \eqref{asymp_infinite}), and the convergent part is of the form
\begin{equation}
\label{fnlrsl}
I_{(c, \Delta)}(\tau) = \frac{\sqrt{-i \tau}}{\sqrt{2\pi}}  \, e^{2\pi i (\Delta - c/24) \tau}.
\end{equation}
It matches the Stanford-Witten DH integral:
\begin{equation}
\label{mtchsw}
I^{\rm DH}_{-(\Delta - c/24)}(- i \tau) =I_{(c, \Delta)}(\tau).
\end{equation}

Note that only for $\Delta>0$ and $b_0=-(\Delta - c/24) < c/24$ the Virasoro coadjoint orbit might give a good quasi-classical description of the corresponding Verma module. Indeed, the weights in the Verma module are always bounded from below by $\Delta$.  For $b_0 < c/24$, the values of the moment map $\mu(b)$ are bounded from above by the $2\pi b_0$ and the image of the coadjoint orbit under the moment map coincides with the convex hull of the weight diagram of the Verma module. For $b_0 > c/24$, this correspondence breaks down. However, equation \eqref{mtchsw} is still valid, and the formal DH integral still determines the asymptotic behaviour of the character.

It is natural to conjecture that for $b_0>c/24$ coadjoint orbits should be replaced by some other symplectic spaces in the orbit - representation correspondence. In particular, in terms of local coordinates $f_n$ on the coadjoint orbit (see \eqref{quadraticform}) one can naturally think of the Lefschetz thimble which corresponds to the $S^1$-fixed point (for general theory, see \cite{Wi2}). Locally, that would mean keeping the reality condition $f_{-n}=\bar{f}_n$ for $n$'s with $n^2> 24b_0/c$ and replacing it with $f_{-n} = - \bar{f}_n$ for a finite number of $n$'s with $n^2 < 24b_0/c$. On such a Lefschetz thimble, $\mu(b)$ is stil a moment map for the $S^1$-action, but now the fixed point is a maximum independently of the value of $b_0$. The formal DH integral remains unchanged and the image of the moment map matches the weight diagram of the representation. We hope to return to this discussion elsewhere.


%
%

\subsubsection{Verma modules with one singular vector}
Next, we consider the case of  Verma modules $V_{(c, \Delta)}$ with one singular vector. The standard parametrisation (borrowed from the Liouville theory) is as follows:
\begin{equation}
\label{cbb}
c= 1 + 6(b + b^{-1})^2, \hskip 0.3cm \Delta= \frac{1}{4}( (b+b^{-1})^2 - (rb + sb^{-1})^2).
\end{equation}
The singular vector is at the level $m=rs$. The character of the irreducible representation $L_{(c,\Delta)}$ reads
\begin{equation}
\label{chione}
\chi_{(c, \Delta)}(\tau) = \frac{q^{\Delta-c/24}(1-q^m)}{\prod_{n=1}^\infty (1 - q^n)} .
\end{equation}
In terms of parameters $b, r, s$ the exponent in the numerator acquires the form
\begin{equation}
\label{dltab}
\Delta - \frac{c}{24} = -\frac{1}{24}-\frac{1}{4} (rb+sb^{-1})^2 = - \frac{1}{24} -\frac{1}{4}(r^2 b^2 + 2rs + s^2 b^{-2}).
\end{equation}
Since we now have several parameters, it opens different possibilities for scaling. If we keep $r$ and $s$ (and, hence, $m=rs$) fixed, then one possible  scaling looks as follows:
\begin{equation}
\label{sclb}
b \mapsto \frac{b}{\sqrt{k}}, \hskip 0.3cm b^{-1} \mapsto \sqrt{k} b^{-1}, \hskip 0.3cm \tau \mapsto \frac{\tau}{k}.
\end{equation}
Under this scaling, the central charge $c(k)$ and the anomalous dimension $\Delta(k)$ behave as follows:
\begin{equation}
\label{sclc}
c(k) = 6b^{-2} k + \dots, \hskip 0.3cm \Delta(k) = \frac{b^{-2}}{4} (1-s^2) k + \dots,
\end{equation}
where $\dots$ stand for subleading terms in $k$. The asymptotic expansion of the character reads
\begin{equation}
\label{twosing}
\chi_{(c(k), \Delta(k))}\left( \frac{\tau}{k}\right) \sim_{k\to \infty} \frac{(2\pi)^{3/2} e^{\frac{\pi i k}{12 \tau}}}{k^{3/2}} \cdot
   \frac{ rs (-i \tau)^{3/2}  e^{- \pi i \tau s^2 b^{-2}/2 } }{\sqrt{2\pi}} = 
   \left(\frac{2\pi}{k}\right)^{3/2} e^{\frac{\pi i k}{12 \tau}} 
    \cdot I_{r,s}(\tau).
\end{equation}
The first factor on the right hand side represents the divergent contribution with $d=-3/2$ and $\alpha=1$, and the second factor is a convergent expression of Duistermaat-Heckman type. For $r=1$, it matches the Stanford-Witten integral:
\begin{equation}
\label{swonen}
I_{1, s}(\tau) = \frac{ s (-i \tau)^{3/2}  e^{- \pi i \tau s^2 b^{-2}/2 } }{\sqrt{2\pi}} = \frac{ s t^{3/2}  e^{\frac{\pi  s^2 (6 b^{-2}) t}{12} }}{\sqrt{2\pi}} = I^{\rm DH}_s(t),
\end{equation}
where $t= -i \tau$ and $6b^{-2} = \lim_{k \to \infty} c(k)/k$.


Note the following interesting geometric phenomenon: the Verma module has a singular vector and the numerator of the character formula has two terms. However, in the asymptotic expansion there is only one exponential factor. This is related to the fact that the highest weight $(c, \Delta)$ and the level $m$ of the singular vector now scale in two different manners (the highest weight scales as $k$ while the level of the singular vector does not). And indeed, Virasoro coadjoint orbits have at most  one $S^1$-fixed point!

Similar to the case of irreducible Verma modules, the coadjoint orbit gives a faithful quasi-classical picture of the Virasoro representation only for $s=1$ (which was actually the case considered in \cite{SW}). In this case, the moment map is bounded from above and its image coincides with the convex hull of the weight diagram. For $s>1$, the moment map is undounded. Nevertheless, equation \eqref{swonen} still holds true and the formal DH integral represents the leading part in the asymptotic expansion of the Virasoro character. Again, it is natural to conjecture that replacing a coadjoint orbit with the Lefschetz thimble of the fixed point would fix the discrepancy between the image of the moment map and the weight diagram of the representation.

We can also consider a more natural scaling:
\begin{equation}
\label{ntrscl}
b \mapsto \frac{b}{\sqrt{k}}, \hskip 0.3cm b^{-1} \mapsto \sqrt{k} b^{-1}, \hskip 0.3cm
 r \mapsto kr, \hskip 0.3cm s \mapsto s, \hskip 0.3cm \tau \mapsto \frac{\tau}{k} .
\end{equation}
In this case, the leading $k$ terms in  both  $c$ and $\Delta$ are proportional to $k$. Furthermore, the level of the singular vector $m=rs$ also scales as $k$.
The asymptotic behaviour of the character is as follows:
\begin{equation}
\label{ntrlchr}
\chi_{(c(k), \Delta(k))}\left(\frac{\tau}{k}\right) \sim_{k\to \infty} 
\left(\frac{2\pi}{k}\right)^{1/2} \exp\left(\frac{\pi i k}{12 \tau}\right) 
\,
\left( \frac{\sqrt{-i\tau}}{\sqrt{2\pi}}  e^{-i \pi (r^2 b^2 + 2rs + s^2 b^{-2})\tau/2} (1 - e^{2\pi i rs \tau}) \right).
\end{equation}

The expression in the parenthesis looks like a sum of two localization contributions coming from two different fixed points. However, there are no Virasoro coadjoint orbits with such geometry.  Hence, it is natural to conjecture that there is a new symplectic space which  which gives rise to a formal DH integral of the type
\begin{equation} \label{two_terms}
I^{\rm DH}_{r,s}(t) = \frac{t^{1/2}}{\sqrt{2\pi}} e^{\pi (r^2 b^2 + 2rs + s^2 b^{-2})t/2} (1 - e^{-2\pi  rs t}).
\end{equation}
It should necessarily have two $S^1$ fixed points, and it is natural to guess that the linear operator induced by the action on the tangent space to each of the fixed points is of the form $A=t \partial_s$, where $t=- i \tau$.

\subsubsection{Verma modules with infinitely many singular vectors}
Verma modules of the Virasoro algebra with infinitely many singular vectors arise when $b^2=p/q$ is a rational number and $1 \leq r \leq q, 1 \leq s \leq p$. It is convenient to introduce notation 
\begin{equation}
\label{mmdl}
P_\pm= r b \pm s b^{-1}.
\end{equation}
Then, the character of the irreducible representation acquires the form
\begin{equation}
\label{chrmmdl}
\chi_{r,s}(\tau) = \frac{\sum_{n \in \mathbb{Z}} (e^{2\pi i \tau (P_++n\sqrt{pq})^2} - e^{2\pi i \tau (P_-+n \sqrt{pq})^2})}{\prod_{n=1}^\infty (1-e^{2\pi i n \tau}) }.
\end{equation}
Again, there are several ways to introduce scaling. The simplest one is as follows:
\begin{equation}
\label{nwsclng}
p \mapsto p, \hskip 0.3cm q \mapsto kq, \hskip 0.3cm r \mapsto kr, \hskip 0.3cm s \mapsto s, \hskip 0.3cm
\tau \mapsto \frac{\tau}{k}.
\end{equation}
Under this scaling, the expressions $(P_\pm +n\sqrt{pq})^2$ scale as $k$, and all exponential factors in the numerator of the character survive without change. The denominator scales as in the case of Verma modules with one or no singular vectors. This yields,
\begin{equation}
\label{nwsclchrct}
\chi_{r, ks}\left(\frac{\tau}{k}\right) \sim 
\left(\frac{2\pi}{k}\right)^{1/2} \exp\left(\frac{\pi i k}{12 \tau}\right) 
 \cdot 
 \frac{(-i\tau)^{1/2}}{\sqrt{2\pi}} \sum_{n \in \mathbb{Z}} (e^{2\pi i \tau (P_++n\sqrt{pq})^2} - e^{2\pi i \tau (P_-+n \sqrt{pq})^2}).
\end{equation}
The finite part of the right hand side looks like a Duistermaat-Heckman type formula with an infinite number of fixed points. Again, as in the case of Verma modules with one singular vector, there are no coadjoint orbits with such geometry and one can conjecture existence of a symplectic space with a formal DH integral of the form
\begin{equation}
\label{nwdhfrm}
I^{\rm DH}_{r,s}(t) = 
 \frac{t^{1/2}}{\sqrt{2\pi}}
 \sum_{n \in \mathbb{Z}} (e^{- 2\pi  (P_++n\sqrt{pq})^2 t} - e^{- 2\pi  (P_-+n \sqrt{pq})^2 t})
\end{equation}

\begin{rem}
One possible candidate for a Hamiltonian ${\rm Diff}(S^1)$-space with an infinite number of fixed points (of the $S^1$-action) has been discussed in \cite{VY}.
\end{rem}

\subsection{Virasoro volume functions}

In this Section, we consider volume functions for Virasoro orbital integrals and for asymptotic expressions for Virasoro characters.

\subsubsection{Formal volume functions}

First, we start with the Stanford-Witten integral $I^{\rm DH}_{b_0}(t)$, and we define the corresponding volume function as the Fourier transform
\begin{equation}
\label{vbzeronw}
V_{b_0}(x) = \int_{i \mathbb{R} + \epsilon} I^{\rm DH}_{b_0}(t) e^{- xt} dt.
\end{equation}
Here the integration is along the imaginary axis slightly shifted in the real positive direction. It is convenient to rewrite this integral in terms of the variable $\tau=i t$:
\begin{equation}
\label{ttrep}
V_{b_0}(x) = - i \int_{\mathbb{R} + i \epsilon} I^{\rm DH}_{b_0}(-i\tau) e^{ix\tau} d\tau =
i \int_{\mathbb{R} + i \epsilon} \frac{(-i \tau)^{1/2} }{\sqrt{2\pi} } \, e^{i(x - 2\pi b_0) \tau } d\tau.
\end{equation}
Note that the integral on the right hand side vanishes if $x > 2\pi b_0$ since the integration contour can be contracted in the upper half plane. For $x < 2\pi b_0$, it is given by the following expression
\begin{equation}
\label{fnlvbz}
V_{b_0}(x) = \frac{2}{\sqrt{2\pi}} \int_0^\infty s^{1/2} e^{-(2\pi b_0 -x) s} ds = 
\frac{1}{ (2\pi b_0 -x)^{3/2} }.
\end{equation}
Vanishing of $V_{b_0}(x)$ for $x > 2\pi b_0$ indicates that $2\pi b_0$ must be the maximal value of the moment map on the orbit. This is only true for $b_0 < c/24$, but the volume function vanishes for higher values of $b_0$ as well. The divergence of $V(x)$ at $x = 2\pi b_0 $ has no clear geometric interpretation (see the next sections for more details).

For the formal DH integral \eqref{frmlans}, a similar computation yields the volume function with a different scaling:
\begin{equation}
\tilde{V}_{b_0}(x) = \frac{C}{(2\pi b_0 -x)^{5/2}}.
\end{equation}

We also test this procedure on the DH type expression \eqref{two_terms} which can be re-written as follows:
\begin{equation}
I^{\rm DH}_{b_0, b_1}(t) = \frac{t^{1/2}}{\sqrt{2\pi}} \, \left( e^{2\pi b_0 t} - e^{2\pi b_1 t} \right),
\end{equation}
where $b_1 = b_0 -m$. The volume function reads
\begin{equation}
V_{b_0, b_1}(x) = \frac{\chi_{(- \infty, 2\pi b_0)}(x)}{(2\pi b_0 - x)^{3/2}} -  \frac{\chi_{(- \infty, 2\pi b_1)}(x)}{(2\pi b_1 - x)^{3/2}},
\end{equation}
were $\chi_E(x)$ is the characteristic function of the set $E \subset \mathbb{R}$.
In addition to singularities at $x \to 2\pi b_{0,1}$, this expression  is negative for $x < 2\pi b_1$. This poses difficulty in interpreting it as a volume function. 
We discuss other possible approaches in the following sections.

\subsubsection{Fourier transforms of character asymptotics}

For calculations of this Section, we recall the following fact:
\begin{equation}   \label{Bessel}
\int_\Gamma  \frac{dt}{t^{\nu+1}} \, e^{\left( At -\frac{B}{t}\right)} = 
\left\{
\begin{array}{ll}
2\pi i \left( \frac{A}{B} \right)^{\frac{\nu}{2}} J_\nu(2\sqrt{AB}) & {\rm if} \,\, A>0, \\
0 & {\rm if} \,\, A<0.
\end{array}
\right.
\end{equation}
Here we choose the branch cut of the function $t^{-(\nu+1)}$ along the negative real axis, the contour $\Gamma$ is the imaginary axis, and it encircles the origin on the right (in the positive real half plane). For $A>0$, this is an instance of the Schl\"afli integral for Bessel functions. For $A<0$, the contour can be contracted in the half-plane ${\rm Re}(t)>0$ and the integral vanishes.

We are now computing Fourier transforms of  asymptotic formulas of Virasoro characters. In contrast to formal DH integrals, they include explicit dependence on the large scaling factor $k$.
It is convenient to use the shorthand notation
\begin{equation}
h=2\pi b_0 - x, \hskip 0.3cm z= \left( \frac{\pi k (2\pi b_0 -x)}{3} \right)^{\frac{1}{2}}.
\end{equation}

We start with the case of irreducible Verma modules and consider the Fourier transform of the right hand side of equation \eqref{lrgscl}:
\begin{equation} \label{V_answer}
V_{b_0}(k, x) = k^{- \frac{1}{2}} \int (- i \tau)^{\frac{1}{2}} \exp\left(\frac{\pi i k}{12\tau} -i  h \tau\right) d\tau = C k^{\frac{1}{2}} h^{-\frac{1}{2}} \left( \frac{\cosh(z)}{z^2} + \frac{\sinh(z)}{z} \right).
\end{equation}
The expression above is valid for $x < 2\pi b_0$. The trigonometric function on the right hand side is the spherical Bessel function $y_1(iz)$. The integral vanishes if $x > 2\pi b_0$.

Note that for $z$ small (this is an artificial regime where $x \to 2\pi b_0$ and $k$ is fixed), the term $\cosh(z)/z^2$ dominates and we recover the behavior $h^{-\frac{3}{2}}$ of the previous Section.
For $k$ large, the logarithm of  $V_{b_0}(k, x)$ has the following interesting asymptotic form
\begin{equation}
\log\left( V_{b_0}(k,x)\right) = u(x) k^{\frac{1}{2}} + v(x) \ln(k) + \dots,
\end{equation}
where $\dots$ stand for subleading terms in $k$. The exponents $u(x)$ and $v(x)$ are as follows 
\begin{equation}  \label{exponents}
u(x) =  \left( \frac{\pi (2\pi b_0 - x)}{3} \right)^{\frac{1}{2}}, \hskip 0.3cm v(x) = 0.
\end{equation}

\begin{conj}
Exponents $u(x)$ and $v(x)$  (and eventually other terms in the asymptotic expansion) are new invariants of reduced spaces $M_x$. 
\end{conj}

These exponents might be suitable infinite dimensional replacements of symplectic volume, but their precise geometric meaning is unknown.
The next example concerns exceptional orbits described by the asymptotic formula \eqref{twosing}. The corresponding Fourier transform is given by
\begin{equation} \label{tildeV}
\tilde{V}_{b_0}(k, x) = k^{- \frac{3}{2}} \int (- i \tau)^{\frac{3}{2}} e^{\frac{\pi i k}{12\tau} +  (2\pi b_0 - x) (- i \tau)} d\tau = C h^{-1} \left( \left(1 + \frac{3}{z^2}\right) \frac{\cosh(z)}{z} + 3 \frac{\sinh(z)}{z^2} \right).
\end{equation}
On the right hand side we have the spherical Bessel function $y_2(iz)$.
For $z$ small, the leading term in $h$ is $h^{-\frac{5}{2}}$ which agrees with the previous Section. For $k$ large, we obtain 
the same leading exponent $u(x)$ as in \eqref{exponents} while the exponent $v(x)=-1/2$ distinguishes the behaviour of $V_{b_0}(k, x)$ and $\tilde{V}_{b_0}(k,x)$.

Finally, we consider the Fourier transform of the expression \eqref{ntrlchr} to obtain
\begin{equation}
V_{b_0, b_1}(k, x) = V_{b_0}(k, x) - V_{b_1}(k,x).
\end{equation}
Here the exponents $u(x)$ and $v(x)$, and all the asymptotic expansion, are the same as in \eqref{V_answer} and \eqref{exponents}. The new term $(-V_{b_1}(k,x))$ is exponentially subleading with respect to $V_{b_0}(k,x)$ and one needs finer methods to detect its influence on the answer.

\subsubsection{Spectral representation and the dual Fourier transfom}

Following \cite{SW} and \cite{SSS}, one can be interested in the spectral representation of formal DH integrals. In more detail, one writes
\begin{equation}
I^{\rm DH}\left( \frac{1}{\beta} \right) = \int_0^\infty \rho(E) e^{-\beta E} \, dE.
\end{equation}
The interpretation of this formula is as follows: one interprets the parameter on the circle as time of a quantum system with the density of states $\rho(E)$. The formal DH integral then plays the role of a partition function of the system. 

One can recover the function $\rho(E)$ by the inverse Laplace transform:
\begin{equation}
\rho(E) = \frac{1}{2\pi} \int I^{\rm DH}\left( \frac{1}{\beta} \right) e^{\beta E} \, d\beta.
\end{equation}
It is surprizing that for Virasoro coadjoint orbits we again obtain Schl\"afli integrals giving rise to spherical Bessel functions. It is convenient to use the notation
\begin{equation}
w = 2(2\pi  b_0 E)^{\frac{1}{2}}.
\end{equation}
For the formal DH integral \eqref{dvir}, we obtain the following spectral representation:
\begin{equation} \label{density1}
\rho_{b_0}(E) = C \int \beta^{-\frac{1}{2}} e^{\left( \frac{2\pi b_0}{\beta} + \beta E \right)} \, d \beta = C (2\pi b_0)^{\frac{1}{2}} \, \frac{\cosh(w)}{w} = C \, \frac{\cosh(2\sqrt{2\pi b_0E})}{\sqrt{E}},
\end{equation}
and for the formal DH integral \eqref{frmlans} we have
\begin{equation}  \label{density2}
\rho_{m}(E)=C \int \beta^{-\frac{3}{2}} e^{\left( \frac{2\pi b_0}{\beta} + \beta E \right)} \, d \beta = C E^{\frac{1}{2}} \, \frac{\sinh(w)}{w} = C  \sinh(2\sqrt{ 2\pi b_0E}),
\end{equation}
where $b_0=m^2c/24$. Density functions \eqref{density1} and \eqref{density2} find their natural interpretation in the matrix model theory \cite{SSS}. As a consequence of equation \eqref{2DH}, we have the following relation between densities $\rho_{b_0}(E)$ and $\rho_m(E)$:
\begin{equation}
\rho_m(E) = \frac{m}{2\pi} \, \left. \frac{\partial \rho_{b_0}(E)}{\partial b_0} \right|_{b_0 = \frac{m^2 c}{24}}.
\end{equation}

It is interesting to consider spectral densities for the leading asymptotic of Virasosro characters. For the leading term in  \eqref{lrgscl}, we obtain
\begin{equation}  \label{density1k}
\rho_{b_0, k}(E) = C k^{-\frac{1}{2}} \int \beta^{- \frac{1}{2}} e^{\left( \frac{2\pi b_0}{\beta} +  \left(E + \frac{\pi k}{12}\right) \beta \right)} \, d \beta = 
  C k^{-\frac{1}{2}} \, \frac{\cosh(2\sqrt{2\pi b_0\tilde{E}})}{\sqrt{\tilde{E}}},
\end{equation}
where $\tilde{E}$ is the renormalised energy:
\begin{equation}
\tilde{E} = E + \frac{\pi k}{12}.
\end{equation}
It is very interesting that the density \eqref{density1k} is exactly of the same form as the density \eqref{density1}. The effect of taking into account the divergent terms is the global rescaling and an energy shift. These observations are also valid for the density function corresponding to \eqref{twosing}:
\begin{equation}
\rho_{m, k}(E) = C k^{-\frac{3}{2}} \,  \sinh(2\sqrt{ 2\pi b_0 \tilde{E}}),
\end{equation}
where, as before, $b_0=m^2 c/24$.

Finally, we would like to point out an interesting link between volume functions $V(x)$ and density functions $\rho(E)$. They a related by a (version of) the Hankel transform:
\begin{equation}
V(x) = x^{-\frac{1}{2}} \, \int_0^\infty E^{\frac{1}{2}} I_{1}(2\sqrt{xE}) \rho(E) dE,
\end{equation}
Where $I_1(z)$ is the modified Bessel function.

We conclude that the volume function $V(x)$ and the spectral density function $\rho(E)$ contain exactly the same amount of information. It would be interesting to extend this observation to higher genus corrections.

\section{Open questions}        \label{sec:final}

In this final Section we collect some open questions concerning infinite dimensional orbital integrals.

\begin{quest}
For Virasoro coadjoint orbits, define Wiener type integrals similar to  Frenkel integrals for coadjoint orbits of $\widehat{LG}$. For previous work on this issue, see {\em e.g.} \cite{DP}.
\end{quest}

\begin{quest}
Find a good replacement for Virasoro coadjoint orbits for $b_0>c/24$. Lefschetz thimbles of the $S^1$ fixed points might be suitable candidates. Among other things, this requires a good complexification of coadjoint orbits. In this context, recall that the Virasoro algebra over $\mathbb{C}$ does not integrate to a group.
\end{quest}

\begin{quest}
Construct (infinite dimensional) symplectic spaces  with the property that their 
formal DH integrals encode the asymptotic behaviour of Virasoro characters for modules with singular vectors. Answering this question will be an important step
towards the orbit method for Virasoro algebra.
\end{quest}

\begin{quest}
Characters of Virasoro representations are examples of conformal blocks on the torus. 
Find interpretation of the scaling limit of other Virasoro conformal blocks (with classical conformal blocks as leading WKB contributions) in terms of orbital integrals. 
\end{quest}

\begin{quest}
Find a natural interpretation of exponents $u(x)$ and $v(x)$ (and possibly other terms in the asymptotic expansion of the function $V(k,x)$) in terms of symplectic geometry of  infinite dimensional reduced space $M_x$ for Virasoro coadjoint orbits.
\end{quest}

\begin{quest}
Construct volume functions for coadjoint orbits of $\widehat{LG}$ and describe the infinite dimensional counterpart of the wall crossing phenomenon for DH functions.
\end{quest}

\begin{quest}
Characters of Virasoro representations have the leading asymptotics defined by the formal DH integral, and subleading terms in the parameter $k$ which are corrections to this leading term.
Following \cite{SSS}, the spectral density $\rho(E)$ admits higher genus corrections. The leading terms of these two expansions match. Is there a relation between subleading terms?
\end{quest}

\vskip 1cm

\noindent {\small {\bf A.A}:  \hspace{.2 cm}  {\sl Department of Mathematics, University of Geneva, 2-4 rue du Li\`evre, c.p. 64,  1211 Geneva 4, Switzerland \hspace{8 cm}\,}\\
\hphantom{xxxxxx} {\it E-mail address}: {\tt Anton.Alekseev@unige.ch}}

\noindent{\small {\bf S.Sh.}: {\sl  The Hamilton Mathematics Institute and the School of Mathematics, \hspace{8 cm}\,
\hphantom{xxxx}   \hspace{4 mm} Trinity College Dublin, Dublin 2, Ireland
\hspace{8 cm}\,\\
\hphantom{xxxx}   \hspace{3 mm} 
Simons Center for Geometry and Physics, Stony Brook, USA 
 \hspace{8 cm}\,
\vspace{1 mm}
On leave of absence from: \hspace{16 cm}\,
\hphantom{xxxx}   \hspace{3 mm} Kharkevich Institute for Information Transmission Problems, Bolshoy Karetny per. 19, Moscow, Russia
 }\\
\hphantom{xxxxxx} \hspace{3 mm}{\it E-mail address}: {\tt samson@maths.tcd.ie}}

\end{document}